\documentclass[aps,prl,showpacs,preprint]{revtex4}

\usepackage{amssymb,amsmath,graphicx}
\allowdisplaybreaks

\newcommand{\mket}[1]{\vert{#1}\rangle}
\newcommand{\mbra}[1]{\langle{#1}\vert}
\newcommand{\UL}{\underline}

\begin{document}

\title{A Robust Method for Estimating the Lindblad Operators of a Dissipative
Quantum Process from Measurements of the Density Operator at Multiple Time Points%
\vspace{0.5in}}
\author{N. Boulant} \author{T. F. Havel} \author{M. A. Pravia} \author{D. G. Cory}
\affiliation{Department of Nuclear Engineering, MIT, Cambridge, Massachusetts 02139%
\vspace{0.5in}}

\begin{abstract}
We present a robust method for quantum process tomography, which yields a set of
Lindblad operators that optimally fit the measured density operators at a sequence of
time points. The benefits of this method are illustrated using a set of liquid-state
Nuclear Magnetic Resonance (NMR) measurements on a molecule containing two coupled
hydrogen nuclei, which are sufficient to fully determine its relaxation superoperator.
It was found that the complete positivity constraint, which is necessary for the
existence of the Lindblad operators, was also essential for obtaining a robust fit
to the measurements. The general approach taken here promises to be broadly useful
in studying dissipative quantum processes in many of the diverse experimental systems
currently being developed for quantum information processing purposes.
\vspace{0.5in}
\end{abstract}

\maketitle

\section{Introduction}
An important task in designing and building devices capable of quantum
information processing (QIP) is to determine the superoperators that describe
the evolution of their component subsystems from experimental measurements.
This task is commonly known in QIP as \emph{Quantum Process Tomography} (QPT)
\cite{NielsenChuang}.
The superoperators obtained from QPT allow one to identify the dominant
sources of decoherence and focus development efforts on eliminating them,
while precise knowledge of relevant parameters can be used to design
quantum error correcting codes the and/or decoherence-free subsystems
that circumvent their effects \cite{BouwmEkertZeili,NielsenChuang}.
Methods have previously been described by which the ``superpropagator''
$\mathcal P$ of a quantum process can be determined \cite{Childs,Fujiwara}.
Assuming that the process' statistics are stationary and Markovian \cite{Alicki,Weiss},
a more complete description may be obtained by determining the corresponding
``supergenerator'', that is, a time-independent superoperator $\mathcal G$
from which the superpropagator at any time $t$ is obtained by solving the
differential equation $\dot{\mathcal P}(t) = -\mathcal G\, \mathcal P(t)$.
The formal solution to this equation is $\exp(-\mathcal Gt)$,
where ``$\exp$'' is the operator exponential.

The purpose of this paper is to describe a data fitting procedure
by which estimates of a supergenerator can be obtained.
This problem is nontrivial because, as in many other data
fitting problems, the estimates obtained from straightforward
(e.g.~least-squares) fits turn out to be very sensitive to small,
and even random, errors in the measured data. In some cases,
this may result in a supergenerator that is obviously physically
impossible; in others, it may simply result in large errors in the
generator despite it yielding a reasonably good fit to the data.
Parameter estimation problems with this property are commonly
known as \emph{ill-conditioned} \cite{RoussLeroy,Tarantola}.
The main result of this paper is that, 
although the problem of estimating a supergenerator from measurements
of the superpropagators at various times is ill-conditioned, this
ill-conditioning can be greatly alleviated by incorporating prior
knowledge of the solution into the fitting procedure as a constraint.
The prior knowledge that we use here is a very general property of
open quantum system dynamics known as \emph{complete positivity}.

Roughly speaking, complete positivity means that if $\mathcal P$ is
a superoperator that maps a density operator of a system to another density operator, then
any extension of the form $\mathcal I\otimes \mathcal P$ also returns
a positive operator, where $\mathcal I$ denotes the identity map on the extension
of the domain.  
The most general form of a completely-positive Markovian master equation
for the density operator $\rho$ of a quantum system is known as
the Lindblad equation \cite{Alicki,Weiss}. This may be written as
\begin{equation}
\dot\rho(t) ~=~ -\imath[ H,\, \rho(t) ] \,+\, \tfrac12\, \sum_{m=1}^M\,
\big( [ L_m\,\rho(t),\;L_m^\dag ] \,+\, [ L_m,\;\rho(t)\,L_m^\dag ] \big) ~,
\label{eq:lind1}
\end{equation}
where $\hbar = 1$, $t$ is time, $H$ is the system's Hamiltonian, the $L_m$
are known as Lindblad operators, and the $L_m^\dag$ denote their adjoints.
It is easily seen that the Lindblad equation necessarily
preserves the trace $\mathrm{tr}(\rho) = 1$ of the
density matrix, meaning ~$\mathrm{tr}(\dot\rho) = 0$,
and a little harder to show that it also preserves the
positive-semidefinite character of the density operator.
Proofs that it has the yet-stronger
property of complete positivity may be found in
Refs.~\cite{Alicki,Lindblad,GorKosSud,HavelPre}.
The QPT method we describe in this paper relies upon
the Lindblad characterization of complete positivity.

The paper is organized as follows.
In the first part of the paper we present a computational procedure
which fits a completely positive supergenerator to a sequence of estimates
of the superpropagators of a quantum process at multiple time points.
This procedure initially extracts an estimate of the decoherent part of
the supergenerator, without the Hamiltonian commutation superoperator
(which is assumed to be available from independent prior measurements).
It then refines this initial estimate via a nonlinear least-squares
fit to the superpropagators, in which complete positivity is enforced
by adding a suitable penalty function to the sum of squares minimized.
Finally, any residual non-completely-positive part of the
supergenerator is ``filtered'' out by a matrix projection technique
based on principle component analysis \cite{HavelPre,Jollife}.

In the second part of the paper, the procedure is validated by using it to 
determine the natural spin-relaxation superoperator of a molecule containing 
two coupled spin $1/2$ nuclei in the liquid state from a temporal sequence
of density operators. These in turn were derived by state tomography,
meaning a set of NMR measurements sufficient to determine the density matrix.
In the process we confirm the ill-conditioned nature of the problem,
and that the  complete positivity constraint is
needed to obtain a robust estimate of the supergenerator.
The final results are used to compute the corresponding Lindblad operators, but these
were difficult to interpret. Hence the
Hadamard representation of $T_2$-relaxation dynamics \cite{HavHad} was used to derive 
a new set of Lindblad operators which are easier to interpret and explain most of the
relaxation dynamics. 
The information these operators convey agree with theoretical
expectations as well as with some additional independent measurements,
in support of the accuracy of the results obtained.

\section{Computational procedure}
In this paper we are concerned with an $N$-dimensional open quantum system ($N <
\infty$), the dynamics of which are described by a Markovian master equation of the
form \cite{Ernst,Weiss,Alicki}:
\begin{equation}\begin{split}
\frac{\mathrm d\rho}{\mathrm dt} = &
-\imath\, [H,\,\rho\,] \,-\, \mathcal R \, (\rho - \rho_\textsf{eq} )
\\[3pt] \Leftrightarrow ~
\frac{\mathrm d\rho_\Delta}{\mathrm dt} = &
-\imath\, \mathcal{H}\,\rho_\Delta \,-\, {\mathcal R}\,\rho_\Delta
\end{split}\end{equation}
In this equation, $\hbar \equiv 1$, $\rho = \rho(t)$ is the system's density operator,
$\rho_\textsf{eq}$ this operator at thermal equilibrium, $\rho_\Delta \equiv \rho -
\rho_\textsf{eq\,}$, $H$ is the system's internal Hamiltonian, $\mathcal{H}$ the
corresponding commutation superoperator, and $\mathcal R$ is the so-called
\emph{relaxation superoperator}. The equivalence of the first and second lines follows
from the fact that $\rho_\textsf{eq}$ is time-independent and proportional to the
Boltzman operator $\exp(-H/k_{\textsf{B}} T )$, so that it commutes with $H$.

By choosing a basis for the ``Liouville space'' of density operators, the equation may
be written in matrix form as \cite{Ernst,HornJohn,HavelPre}
\begin{equation}
\frac{\mathrm d \mket{\UL{\,\smash\rho}_\Delta}}{\mathrm d\hspace{0.05em}t} ~=~
-\big(\imath\,\UL{\mathcal{H}} \,+\, \UL{{\mathcal R}}\, \big) \mket{\UL{\,\smash\rho}_\Delta}
~\equiv~ -\UL{\mathcal G}\, \mket{\UL{\,\smash\rho}_\Delta}
\end{equation}
where the underlines denote the corresponding matrices and $\mket{\UL{\,\smash\rho}}$
is the $N^2$-dimensional column vector obtained by stacking the columns of the
density matrix $\UL{\,\smash\rho}$ on top of each other in left-to-right order\cite{HavelPre}.
A numerical solution to this equation at any point $t$ in time may be obtained
by applying the propagator $\UL{\mathcal{P}}(t)$ to the initial condition
$\mket{\UL{\,\smash\rho}_\Delta(0)}$, where the propagator is obtained
by computing the matrix exponential $\UL{\mathcal{P}}(t) =
\UL{\smash\exp} \big(\!-\! \UL{\mathcal G}\, t\hspace{0.05em}
\big)$ \cite{HornJohn,NajHav}.
Note that $\UL{\mathcal{H}}$ and $\UL{\mathcal R}$ do not commute in general,
and that the sum $\UL{\mathcal G} \equiv\imath\, \UL{\mathcal{H}} + \UL{\mathcal R}$
will not usually be a normal matrix (one which commutes with its adjoint).
This in turn greatly reduces the efficiency and stability with
which its matrix exponential can be computed \cite{Moler+vanLoan}
(although this was not an issue in the small problems dealt with here),
and we expect it to also significantly complicate the \emph{inverse problem}.

In this section we describe an algorithm for solving this inverse problem, that is
to determine the relaxation superoperator $\UL{\mathcal R}$ from an estimate of the
Hamiltonian $\UL H$ together with estimates of the propagator $\UL{\mathcal{P}\!}_{\,m}
= \UL{\mathcal{P}}(t_m)$ at one or more time points $t_1<t_2< \ldots < t_M$.
This problem, like many other inverse problems, turns out to be ill-conditioned,
meaning that small experimental errors in the estimates of the $\UL H$ and
$\UL{\mathcal{P}\!}_{\,m}$ may be amplified to surprisingly large, and generally
nonphysical, errors in the resulting superoperator $\UL{\mathcal R}$ \cite{RoussLeroy}.
For example, if one tries to estimate $\UL{\mathcal R}$ in the obvious way as
\begin{equation}
\UL{\mathcal R} ~\approx~ \Big( -\imath\, \UL{\mathcal{H}} \,-\,
\UL{\smash\log}\big(  \UL{\mathcal{P}\!}_{\,1} \big) \Big) \Big/ t_1 ~,
\end{equation}
one will generally obtain nonsense even if $\UL{\mathcal{H}}$
and $\UL{\mathcal{P}\!}_{\,1}$ are known to machine precision,
because of the well-known ambiguity of the matrix logarithm
with respect to the addition of independent multiples of
$2\,\imath\,\pi$ onto its eigenvalues.
Using the principal branch of the logarithms will only
work if $\mathcal{H}$ is small compared to ${\mathcal R}$,
and the only reasonably reliable means of resolving the ambiguities
is to utilize additional data and/or prior knowledge of the solution.
Even then, a combinatorial search for the right multiples
of $2\,\imath\,\pi$ may be infeasibly time-consuming.

An alternative to the logarithm which utilizes data at multiple time points and is
capable of resolving the ambiguities even when $\mathcal{H}$ is much larger than
${\mathcal R}$ is to estimate the derivative at $t = 0$ of
\begin{equation}
\UL e^{it\UL{\mathcal{H}}/2}\, \UL{\mathcal{P}}(t)\, \UL e^{it\UL{\mathcal{H}}/2}
~=~ \UL e^{-t\,\UL{\mathcal R}} \,+\, O(t^2) ~.
\end{equation}
This derivative is obtained by Richardson extrapolation using central differencing
about $t = 0$ over a sequence of time points such that $t_m = 2^{m-1} t_1$ ($m =
1,\ldots,M$), according to the well-known procedure \cite{DahlBjor}: \pagebreak[3]
\begin{tabbing}
\hspace{18pt} \= \hspace{18pt} \= \hspace{18pt} \= \hspace{18pt} \kill
\> \texttt{for $m$ from $1$ to $M$ do} \\
\>\> $\UL{\mathcal{D}}_{\,1,1+M-m} ~:=~ 2^{m-2\,}\big( \UL e^{it_m \UL{\mathcal{H}}/2}\, \UL{\mathcal{P}\!}_{\,m}\, \UL e^{it_m \UL{\mathcal{H}}/2} \,-\, \UL e^{-it_m \UL{\mathcal{H}}/2}\, \UL{\mathcal{P}\!}_{\,m}^{\;-1}\, \UL e^{-it_m \UL{\mathcal{H}}/2} \big)\,$; \\
\> \> \texttt{for $\ell$ from $1$ to $m-1$ do} \\
\> \> \> $\UL{\mathcal{D}}_{\,1+\ell,1+M-m} ~:=~ \UL{\mathcal{D}}_{\,\ell,1+M-m} \,+\, (\UL{\mathcal{D}}_{\,\ell,1+M-m} \,-\, \UL{\mathcal{D}}_{\,\ell,M-m}) / (4^\ell - 1)\,$; \\
\> \> \texttt{end do} \\
\> \texttt{end do}
\end{tabbing}
The inverse $\UL{\mathcal{P}\!}_{\,m}^{\;-1} = \UL{\mathcal P}(-t_m)$ is assured
of existing unless long times are used or the errors in the data are large.
The method produces an estimate of the
derivative at $t = 0$ that is accurate up to $O(t_{\,1}^{\,2M})$, and which may be
increased by computing the exponential from the highest-order estimate at further interval
halvings. The method performs well when the relative errors in the Hamiltonian $\delta
\ll 1/(\Delta\nu\,t_M)$, where $\Delta\nu$ is the range of frequencies present in the
Hamiltonian, but it tends to emphasize the errors in $\UL{\mathcal{P}\!}_{\,1}$ rather
than averaging over the errors at all the time points. Hence we do not recommend
that it be used alone, but rather as a means of obtaining a good starting point
for a nonlinear fit to the data, as will now be described.

This nonlinear fit involves minimization of the sum of squares
\begin{equation}
\chi^2\big( \UL{\mathcal R};\, \UL{\mathcal{H}}, \UL{\mathcal{P}\!}_{\,1\,}, \ldots, \UL{\mathcal{P}\!}_{\,M}
\big) ~\equiv~ \sum_{m=1}^M\, \Big\|\, \UL{\smash\exp}\big( -\!(\imath\, \UL{\mathcal{H}} +
\UL{\mathcal R})\, t_m\, \big) \;-\; \UL{\mathcal{P}\!}_{\,m} \Big\|^{2}_\textsf{F}
\end{equation}
with respect to $\UL{\mathcal R}$, where  $\|\cdot\|^{2}_\textsf{F}$ denotes the squared
Frobenius norm (sum of squares of the entries of its matrix argument). Previous
results with similar minimization problems indicate that $\chi^2$ will have many local
minima \cite{Kwaku}, making a good starting point absolutely necessary (even disregarding
the $2\imath\pi$ ambiguities discussed above). The derivatives of this function may be obtained
via the techniques described in Ref. \cite{NajHav}, but the improvements in efficiency
to be obtained by their use are likely to be of limited value in practice given that
all the resources needed, both experimental and computational, grow rapidly with $N$
(which itself grows exponentially with the number of qubits used in quantum
information processing problems). In addition, the quality of the results matters a
great deal more than the speed with which they are obtained, and the quality will not
generally depend greatly on the accuracy with which the minimum is located.

For these reasons, we have used the Nelder-Mead simplex algorithm \cite{NeldMead},
as implemented in the \textsf{MATLAB}$^\mathsf{TM}$ program, for the small (two qubit)
experimental test problem described in the following section. This has the further
advantage of being able to avoid local minima better than most gradient-based
optimization algorithms. Preliminary numerical studies, however, exhibited the
anticipated ill-conditioning with respect to small perturbations in the data, even
when $\UL{\mathcal R}$ was constrained to be symmetric (implying a unital system
which satisfies detailed balance) and positive semidefinite (as required for the
existence of a finite fixed point). Therefore it is necessary to incorporate
additional prior information regarding $\UL{\mathcal R}$ into the minimization.
The information that we have found to be effective is a property of $\UL{\mathcal R}$
known as \emph{complete positivity} \cite{Weiss,Alicki,HavelPre}.

An \emph{intrinsic} definition which does not involve an environment
was first given by Kraus \cite{Kraus2}, and states that a superpropagator
$\mathcal P$ is completely positive
if and only if it can be written as a ``Kraus operator sum'', namely
\begin{equation}
\mathcal P(\rho) ~=~ \sum_{\ell=1}^{N^2}\, K_{\rule{0pt}{1.45ex}\ell}\,
\rho\, K_\ell^\dag ~,
\end{equation}
where $\rho = \rho^\dag$ and $K_1, \ldots, K_{N^2}$ all act on the system alone.
Another intrinsic definition subsequently given by Choi \cite{Choi}
states that a superpropagator is completely positive if and only if,
relative to any basis of the system's Hilbert space, the so-called
\emph{Choi matrix} is positive semidefinite \cite{HavelPre}, namely
\begin{equation}
\UL{0\rule[-0.3ex]{0pt}{0pt}} ~\preceq~ \UL{C\rule[-0.2ex]{0pt}{0pt}} ~\equiv~
\sum_{i,j=0}^{N-1}\, \UL{\mathcal P\smash{(\mket{\hspace{.5pt}i\hspace{.5pt}}\mbra{j})}}
\otimes (\mket{\hspace{.5pt}i\hspace{.5pt}}\mbra{j}) ~=~ \sum_{i,j=0}^{N-1}\,
\UL{\mathcal P\rule[-0.2ex]{0pt}{0pt}}(\mket{\hspace{.5pt}i\hspace{.5pt}}\mket{j})
(\mbra{\hspace{.5pt}i\hspace{.5pt}}\mbra{j}) ~.
\end{equation}
This equation uses the notation of quantum information processing, in which the $i$-th
elementary unit vector is denoted by $\mket{\hspace{.5pt}i\hspace{.5pt}}$ ($0 \le i < N$),
$\UL{\mathcal P\smash{(\mket{\hspace{.5pt}i\hspace{.5pt}}\mbra{j})}}$ is the $N \times N$
matrix of the operator obtained by applying the superpropagator $\mathcal P$ to the
projection operator given by $\mket{\hspace{.5pt}i\hspace{.5pt}}\mbra{j}$ versus our
choice of basis, and $\UL{\mathcal P\rule[-0.2ex]{0pt}{0pt}}$ is the $N^2 \times N^2$
matrix for $\mathcal P$ versus the Liouville space basis $\mket{\hspace{.5pt}i\hspace{.5pt}}
\mket{j}$ (as for $\UL{\mathcal H}$ etc. above).
It can further be shown that the eigenvectors $\UL k_{\,\ell}$ of the
Choi matrix are related to the (matrices of an equivalent set of) Kraus
operators by $\mket{\UL{K\!}_{\,\ell}} = \sqrt{\kappa_\ell}\, \UL k_{\ell\,}$,
where $\kappa_\ell\ge0$ are the corresponding eigenvalues and the ``ket''
$\mket{\UL{K\!}_{\,\ell}}$ indicates the column vector obtained by stacking
the columns of $\UL{K\!}_{\,\ell}$ on top of one another \cite{HavelPre}.

These results can be used not only to compute a Kraus operator sum from any completely
positive superpropagator given as a ``supermatrix'' acting on the $N^2$-dimensional
Liouville space, but also to ``filter'' such a supermatrix so as to obtain the
supermatrix of the completely positive superpropagator nearest to it, in the sense of
minimizing the Frobenius norm of their difference \cite{HavelPre}. This is done simply
by setting any negative eigenvalues of the Choi matrix to zero, rebuilding it from the
remaining eigenvalues and vectors, and converting this reconstructed Choi matrix back
to the corresponding supermatrix. Although this generally has a beneficial effect upon
the least-squares fits versus $\chi^2$ (as defined above), it is still entirely
possible that the sequence of filtered propagators ${\mathcal P\,}_{\!m}'$ will
not correspond to a completely positive \emph{Markovian} process, so that no
time-independent relaxation superoperator ${\mathcal R}$ can fit it precisely.
This, together with the ill-conditioned nature of the problem, implies one
will still not usually obtain satisfactory results even after filtering.
For this reason we shall now describe how the above characterizations
of completely positive superpropagators can be extended to supergenerators.

As indicated in the Introduction, completely positive Markovian
processes, or \emph{quantum dynamical semigroups} as they are also known,
may be characterized as those with a generator $\mathcal G$ that can be
written in Lindblad form \cite{GorKosSud, Lindblad, Weiss, Alicki, HavelPre}.
On expanding the commutators in Eq~(\ref{eq:lind1}), this becomes
\begin{equation}
-\mathcal G(\, \rho\, ) ~\equiv\, -\imath \big[ H,\,
\rho\, \big] \;+\; \tfrac12\, \sum_{m=1}^{N^2}\, \Big( 2\, L_{m\rule{0pt}{1.4ex}}\,
\rho\, L_m^\dag \,-\, L_m^\dag L_{m\rule{0pt}{1.4ex}}\, \rho \,-\, \rho\, L_m^\dag
L_{m\rule{0pt}{1.4ex}} \Big) ~.
\label{eq:qds}
\end{equation}
The operators $L_m$ are usually called \emph{Lindblads}.
It may be seen that the Choi matrix $\UL{\mathcal C}$ associated
with $-\UL{\mathcal R}$ is \emph{never} positive semidefinite,
because $\mbra{\UL{I}}\, \UL{\mathcal C}\, \mket{\UL{I}}
= \mathrm{tr}(-\UL{\mathcal R}) < 0$.
Nevertheless, it can be shown that any trace-preserving $\UL{\mathcal R}$
(meaning $\mathrm{tr}(\UL{\mathcal R}(\UL{\smash\rho})) = 0$) is completely positive
if and only if a certain projection of $\UL{\mathcal C}$ is positive semidefinite,
namely $\UL{\mathcal E}\, \UL{\mathcal C}\, \UL{\mathcal E}$ where
$\UL{\mathcal E} = \UL I\otimes\UL I-\mket{\UL{I}}\,\mbra{\UL{I}}\;$ \cite{HavelPre}.
In this case an equivalent system of orthogonal Lindblads
is determined by $\mket{\UL{L}_m} ~=~ \sqrt{\lambda_m}\,
\UL\ell_{\,m\,}$, where $\lambda_m \ge0$ are the
eigenvalues and $\UL\ell_m$ the eigenvectors of
$\UL{\mathcal E}\, \UL{\mathcal C}\, \UL{\mathcal E}$.
In the event that $\UL{\mathcal E}\, \UL{\mathcal C}\, \UL{\mathcal E}$
has negative eigenvalues we can simply set them to zero to obtain a similar
but completely positive supergenerator, much as we did with the Kraus operators.
Most importantly, however, this characterization of completely positive
supergenerators gives us a means of enforcing complete positivity during
nonlinear fits to a sequence of propagators at multiple time points.

The following section describes our experience with applying this approach to a
sequence of propagators obtained from liquid-state NMR data. The complete positivity
of the relaxation superoperator was maintained by adding a simple penalty function
onto the sum of squares that was minimized by the simplex algorithm, as described
above. This penalty function consisted of the sum of the squares of the negative
eigenvalues of the corresponding projected Choi matrix. While more rigorous and
efficient methods of enforcing the projected Choi matrix to be positive semidefinite
are certainly possible, this strategy was sufficient to demonstrate
the main result of this paper, which is that \emph{the complete positivity
constraint greatly alleviates the ill-conditioned nature of such fits}.

\section{Experimental Validation}
The experiments were carried out on a two-spin $\frac{1}{2}$ system consisting
of the hydrogen atoms in 2,3-dibromothiophene (see Fig.~1) at $300 K$ dissolved
in deuterated acetone, using a Bruker Avance 300 MHz spectrometer.
\begin{figure}
\includegraphics[width=6cm,height=6cm]{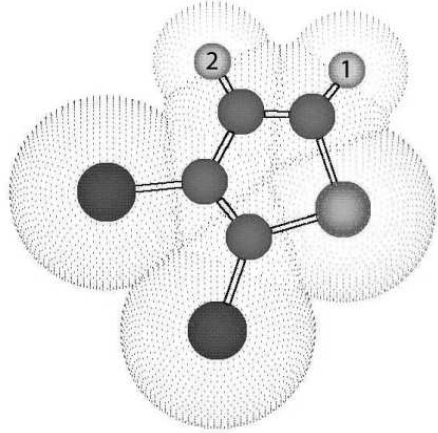}
\caption{\small{Molecule of 2,3-dibromothiophene with the two protons labeled $1$ and $2$.
The chemical bonds among the atoms are indicated by double parallel lines,
and a transparent ``dot-surface'' used to indicate their van der Waals radii.}}
\end{figure}
The internal Hamiltonian of this system in a frame rotating at the frequency of the
second spin is
\begin{eqnarray}
H ~\equiv~ H_{int} ~=~ \pi\, (\, \nu_1\, \sigma_{\textsf z}^1 \;+\;
\tfrac{J}{2}\, \sigma^1 \cdot \sigma^2\, )
\end{eqnarray}
where $\nu_1=161.63$Hz is the chemical shift of the first spin,
$J=5.77$Hz is the coupling between the spins, and
$\sigma = [\sigma_{\textsf x},\, \sigma_{\textsf y}, \sigma_{\textsf z}]$
are Pauli spin operators.

The ``quantum operation'' we characterized was just free-evolution
of the system  under its internal Hamiltonian, together with
decoherence and relaxation back towards the equilibrium state
$\rho_{\textsf eq} \sim\, \sigma_{\textsf z}^1 + \sigma_{\textsf z}^2$.
In liquid-state NMR on small organic molecules like dibromothiophene,
this process is mediated primarily by fluctuating dipolar interactions
between the two protons as well as with 
spins neighboring molecules, and since the correlation time for small
molecules in room temperature liquids is on the order of picoseconds,
the Markovian approximation is certainly valid \cite{Ernst,Freeman}. We add
that our sample was not degassed so that the presence of dissolved paramagnetic
$O_2$ shortened the $T_1$ and $T_2$ relaxation times. 

The experiment consisted of preparing a complete set of orthogonal input states
(that is, density matrices), letting each evolve freely for a given time $T$,
and then determining the full output states via quantum state tomography
\cite{Chuang3,CoryGroup}. Since only ``single quantum'' coherences can be
directly observed in NMR \cite{Ernst,Freeman}, this involves repeating the
experiment several times followed by a different readout pulse sequence each time,
until all the entries of the density matrix have been mapped into observable ones.
The experiments were carried out at four exponentially-spaced times $T$,
as required by the Richardson extrapolation procedure described above,
specifically $T = 0.4,~0.8,~1.6$ and $3.2$s.

\begin{table}
\begin{tabular}{|ccc|ccc|}
\hline
~index~ & ~operator~ & ~order~ & ~index~ & ~operator~ & ~order~ \\
\hline\hline
$1$ & $4\!\times\!4$ identity & $0$ & $9$ & $\sigma_{\textsf x}^2$ & $1$ \\
$2$ & $\sigma_{\textsf z}^1$ & $0$ & $10$ & $\sigma_{\textsf y}^2$ & $1$ \\
$3$ & $\sigma_{\textsf z}^2$ & $0$ & $11$ & $\sigma_{\textsf x}^1\sigma_{\textsf z}^2$ & $1$ \\
$4$ & $\sigma_{\textsf z}^1\sigma_{\textsf z}^2$ & $0$ & $12$ & $\sigma_{\textsf y}^1\sigma_{\textsf z}^2$ & $1$ \\
$5$ & $\sigma_{\textsf x}^1\sigma_{\textsf x}^2+\sigma_{\textsf y}^1\sigma_{\textsf y}^2$ & $0$ & $13$ & $\sigma_{\textsf z}^1\sigma_{\textsf x}^2$ & $1$ \\
$6$ & $\sigma_{\textsf x}^1\sigma_{\textsf y}^2-\sigma_{\textsf y}^2\sigma_{\textsf x}^1$ & $0$ & $14$ & $\sigma_{\textsf z}^1\sigma_{\textsf y}^2$ & $1$ \\
$7$ & $\sigma_{\textsf x}^1$ & $1$ & $15$ & $\sigma_{\textsf x}^1\sigma_{\textsf x}^2-\sigma_{\textsf y}^1\sigma_{\textsf y}^2$ & $2$ \\
$8$ & $\sigma_{\textsf y}^1$ & $1$ & $16$ & $\sigma_{\textsf x}^1\sigma_{\textsf y}^2+\sigma_{\textsf y}^1\sigma_{\textsf x}^2$ & $2$ \\
\hline
\end{tabular}
\caption{~Table of operators (versus Cartesian basis) in the transition basis
used for the density and superoperator matrices, the corresponding matrix indices
and their coherence orders (see text).}
\end{table}

To describe the density and superoperator matrices,
the so-called ``transition basis'' was used \cite{Ernst}.
This Liouville space basis is intermediate between the Cartesian basis
and the Zeeman (or polarization and shift operator \cite{Ernst}) basis,
in that the basis elements are all Hermitian like those of the Cartesian basis,
but like the Zeeman basis they have a well-defined \emph{coherence order},
or difference in total angular momentum along the applied magnetic field
$B_0$ between the two Zeeman states connected by the transition.
These basis states are listed in TABLE I versus the Cartesian basis.

This basis was chosen because the relaxation superoperator ${\mathcal R}$ is
expected to have the ``Redfield kite'' structure in this basis \cite{Ernst}.
This block diagonal structure arises because the difference in frequency
between transitions of different coherence orders, given that the Zeeman
interaction dominates all others, is large enough to average out
these other interactions including those responsible for decoherence
and relaxation, in effect decoupling the blocks from one another
so that no \emph{cross relaxation} occurs between them (see Fig.~2).
\begin{figure}[htb]
\includegraphics[width=8cm,height=7cm]{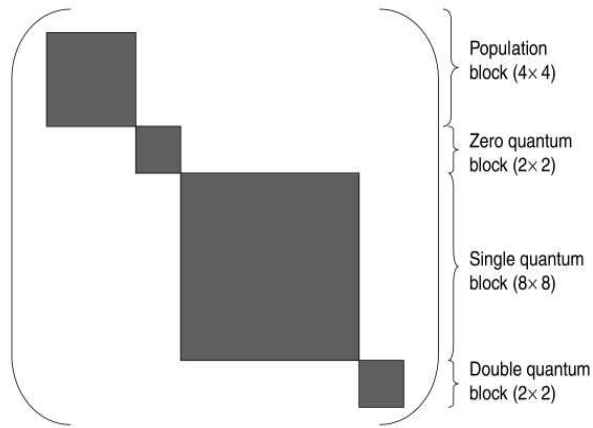}
\caption{\small{Redfield kite struture of the relaxation superoperator expressed
in the transition basis. The non-shaded area corresponds to 
Blocks of different coherence order are decoupled from each other.}}
\end{figure}
This so-called ``secular approximation'' considerably
reduces the number of parameters in the superoperator
from $256 = ((2^2)^2)^2$ to $81 = 3^2 + 2^2 + 8^2 + 2^2$
(since neither the identity nor the other diagonal ($\sigma_{\textsf z}$)
basis elements are expected to cross relax with any non diagonal elements).

An additional reduction may be obtained by assuming detailed balance:
the microscopic reversibility of all cross relaxation processes.
The relaxation superoperator reconstructed from the
experimental data was bordered with an initial row and column of
zeros to force $\mathcal R(\,I\,) = 0$, because the totally random
density matrix $I/4$ cannot be observed by NMR spectroscopy and because
of the trace-preserving process assumption that we made.
This may be done providing $\mathcal R$ operates
on $\rho_\Delta = \rho - \rho_{\mathsf{eq}}$,
and together with detailed balance it implies that
the supermatrix $\UL{\mathcal R}$ will be symmetric,
reducing the number of parameters to be
estimated to only $48 = 6 + 3 + 36 + 3$.

\begin{figure}[htb]
\includegraphics[width=10cm,height=9cm]{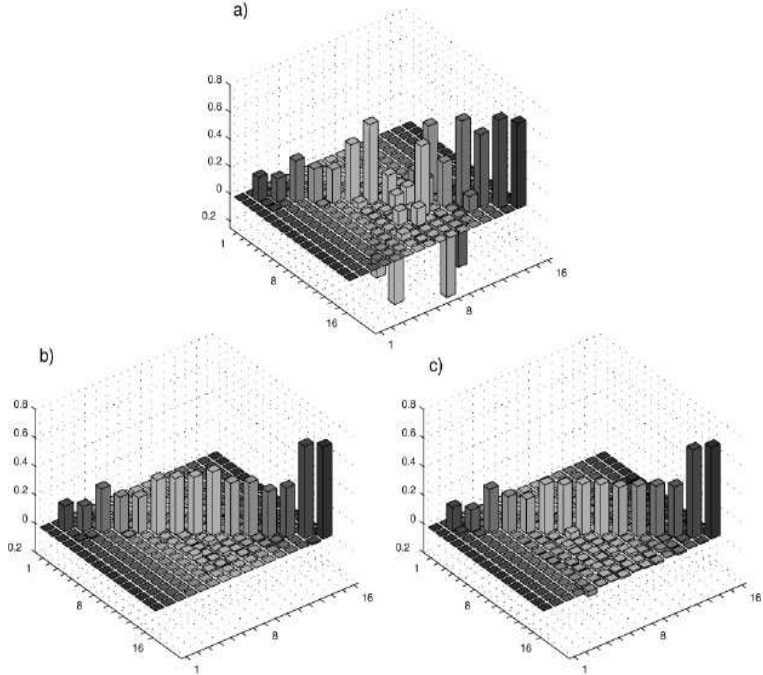}
\caption{\small{Three different estimates of the relaxation superoperator of
$2,3$-dibromothiophene in the transition basis, indexed as indicated in Table I.
\textbf{(a)} Relaxation superoperator obtained from a least-squares fit,
without the complete positivity constraint, of the exponential
$\UL{\smash\exp}(-\imath(\UL{\mathcal H} + \UL{\mathcal R})t_m)$ to
the propagators $\UL{\mathcal P}_m$ at the corresponding times
($t_1 = 0.4, t_2 = 0.8, t_3 = 1.6, t_4 = 3.2$s) with respect
to the unknown relaxation superoperator $\UL{\mathcal R}$,
starting from the results of Richardson extrapolation (see text).
\textbf{(b)} The relaxation superoperator obtained from a fit to
the same data and with the same starting value of $\UL{\mathcal R}$,
but with the complete positivity constraint included in the fit.
\textbf{(c)} The relaxation superoperator obtained by assuming
that $\mathcal H$ and $\mathcal R$ commute, and using the
average of the estimates obtained by taking the logarithms
of the absolute values of the eigenvalues of the propagators
over all four time points as the final estimate (see text).}
\label{fig:results}}
\end{figure}

The result of applying the fitting procedure without the complete
positivity constraint to the initial estimate obtained by
Richardson extrapolation is shown in FIG. \ref{fig:results}(a).
It may be seen that the rates did not vary in
a systematic fashion with the coherence order,
and that large cross-relaxation rates were found,
which is not consistent with the physics of
spin relaxation in liquid-state NMR spectroscopy.
In addition, this relaxation superoperator implies that spin $1$
has a $T_2 \approx 2.3s$  while spin $2$ has a $T_2 \approx 4.6$s,
in disagreement with the independent measurements of $T_2$ given below.

The fit after adding the complete positivity
constraint is shown in FIG. \ref{fig:results}(b),
again starting from the results of Richardson extrapolation.
It may now be seen that the results do vary systematically with coherence number,
and that the resulting relaxation superoperator is very nearly diagonal.
To obtain further evidence for the validity of this superoperator,
we measured the single spin $T_1$ (longitudinal, or $\sigma_{\textsf z}$)
and $T_2$ (transverse, or $\sigma_{\textsf x} = \sigma_{\textsf y}$) relaxation rates, using
the well-established inversion-recovery and CPMG experiments \cite{Freeman}.
The results for both spins were $T_1 = 5.6$s and $T_2 = 2.7$s,
which agree quite well with the values of $5.6 s$ and $2.6 s$ 
obtained from this relaxation superoperator.
Although this is obviously a relatively simple relaxation superoperator,
it is reasonable to expect that a complete positivity constraint will substantially
improve the estimates of more complicated superoperators containing nonzero
cross-relaxation rates that cannot be obtained from standard experiments.

Finally, the cross-relaxation rate between the population
terms $\sigma_{\textsf z}^1$ and $\sigma_{\textsf z}^2$,
which is due to the well-known nuclear Overhauser effect
\cite{Solomon}, is essentially zero in Fig.~3(b).
This can occur when the overall rotational correlation time of
the molecule plus its ``solvent-cage'' is on the order of $1$ns,
but was somewhat unexpected given the small size of $2,3$-dibromothiophene.
As a result, we carried out the selective inversion recovery experiment
that consists of inverting selectively the longitudinal magnetization of
one of the two protons and looking at the evolution of the magnetization
of the other one while the first relaxes towards thermal equilibrium. The
change in longitudinal magnetization of the second proton was measured to be less 
than $1\%$ of the unperturbed magnetization essentially revealing no NOE effect and
providing yet further evidence for the validity of this superoperator. The discrepancy
between this result and the one given in \cite{Freeman} is explained by the presence
of dissolved paramagnetic $O_2$ in our sample so that the $T_1$ relaxation time
was shortened in such a way that the NOE effect became almost unobservable (our $T_1$ was
measured to be $5.6s$ while the $T_1$ reported in \cite{Freeman} is around $47.5s$).

Because of the substantial degeneracy of the diagonal elements with the same
coherence order, the superoperator in Fig.~3(b) was also very nearly diagonal
in the eigenbasis of the Hamiltonian commutation superoperator $\mathcal H$,
so that $\mathcal H$ and $\mathcal R$ very nearly commute. This allowed further
estimates to be obtained directly from the superpropagators $\UL{\mathcal P}_{\,m}
\approx \UL{\smash\exp}(-\imath\UL{\mathcal H}t_m)\, \UL{\smash\exp}(-\UL{\mathcal R}t_m)$,
simply by taking the (real) logarithms of the absolute values of their eigenvalues
and thereby cancelling the phase factors from the Hamiltonian's exponential.
From FIG.~\ref{fig:results}(c) we see that the result of averaging
these estimates over all four evolution times is very similar to
the completely positive estimate in FIG.~\ref{fig:results}(b)
(correlation coefficient $0.80$; squared norms of difference over average $0.90$).
We note that the estimate in FIG.~\ref{fig:results}(c)
did not explicitly assume the Redfield kite structure,
thereby providing a further consistency check on our results.

\section{Interpretation via Lindblad and Hadamard Operators}
In this section we present a system of Lindblad operators which act
on the density operator to give essentially the same derivative as the
relaxation superoperator described above (see FIG.~\ref{fig:results}).
As described in the foregoing ``Computational Procedure'' section,
such a system of Lindblad operators may be obtained by
diagonalizing the corresponding projected Choi matrix,
although it will be seen that a more easily interpreted
system was obtained by considering the parts of $\mathcal R$
responsible for $T_1$ and $T_2$ relaxation separately and by using the
so-called "Hadamard relaxation matrix" formalism \cite{HavHad}. Because these
calculations were somewhat involved, however, the 
details of the Hadamard operators calculations are given in an appendix.
We add that from here on, the relaxation superoperator $\mathcal R$ will refer
to the matrix shown in FIG.~\ref{fig:results}(b).

These representations of relaxation processes are normally applied to the
density matrix in the Zeeman basis $\UL{\smash\rho}_\Delta^{\textsf{zee}}$
(regarded as the computational basis in QIP), which requires converting
the supergenerator $\mathcal R$ from the transition to the Zeeman basis.
This is easily done via a unitary transformation, $\UL{\mathcal R}^{\textsf{zee}}
~\equiv~ 2\, \UL U\, \UL{\mathcal R}^{\textsf{tra}}\, \UL U^\dag$, where
$\UL U\, \mket{\UL{\smash\rho}^{\textsf{tra}}} = \mket{\UL{\smash\rho}^{\textsf{zee}}}$
(the matrix $\UL U$ may be derived from TABLE I; the factor of $2$ corrects
for a change in norm due to the fact that the transition basis is Hermitian).
Although any relaxation superoperator can be modified to act directly on
the density operator $\rho$ rather than its difference with the equilibrium
density operator $\rho_\Delta = \rho - \rho_{\mathsf{eq}}$ (\emph{vide supra})
by taking the right-projection $\UL{\mathcal R}\, \big(
\UL{\mathcal I} \,-\, \mket{\UL{\rho}}_{\mathsf{eq}}\mbra{\UL{\mathcal I}} \big)$ \cite{Levante,Ghose},
this makes only a negligible change to $\mathcal R$
since in liquid-state NMR $\rho_{\textsf{eq}}$ differs
from the identity $I$ by $\lesssim\!10^{-5}$.
For that reason, the treatment of $T_1$ relaxation given below
was considerably simplified by treating it as a unital
(identity preserving) process acting on $\rho_\Delta$.

As described following Eq.~(\ref{eq:qds}), a complete system of Lindblad
operators may be obtained by diagonalizing the projected Choi matrix
\begin{equation}
\UL{\mathcal E}\, \UL{\mathcal C}\, \UL{\mathcal E}
~=~ \UL V\, \UL\Lambda\, \UL V^\dag ~,
\end{equation}
where it is assumed the eigenvalues have been ordered such
that $\lambda_m \ge \lambda_{m+1}$ for $m = 1,\ldots,N^2-1$,
and defining the Lindblad matrices such that for all $\lambda_m > 0$:
\begin{equation}
\mket{\,\UL L_m\,} ~=~ \sqrt{\lambda_m}\, \UL V\, \mket{m} ~.
\end{equation}
This gave rise to a total of $11$ Lindblads, the phases of which
were chosen so as to make them as nearly Hermitian as possible.
Once this was done, all $11$ were within $2$\% of being Hermitian.

The relative contributions of these Lindblads to the overall relaxation
of the spins can be quantified by the squared Frobenius norms $\|\UL L_m
\|_{\textsf F}^2 = \lambda_m$.
This calculation shows that about $35$\% of the mean-square noise resided in the first Lindblad,
\begin{equation}
\UL L_1 ~\approx~ 0.346\, \big( \UL{\sigma}_{\textsf z}^1 + \UL{\sigma}_{\textsf z}^2 \big)
\,+\, 0.025\, \UL{\sigma}_{\textsf z}^1 \UL{\sigma}_{\textsf z}^2 ~,
\end{equation}
which represents strongly correlated dephasing with a $T_2$
for both spins of $\sim4.2$s \cite{HavHad}, much as expected.
The next four largest Lindblads together contributed, roughly
equally, another $43$\% to the total mean square noise,
but were considerably more difficult to interpret:
\begin{align}
\UL L_2 ~\approx~ &
\begin{aligned}[t] &
-0.013\, \UL{\sigma}_{\textsf x}^1 \,-\, 0.045\, \UL{\sigma}_{\textsf y}^2
\,-\, 0.153\, \UL{\sigma}_{\textsf x}^2 \,-\, 0.061\, \UL{\sigma}_{\textsf y}^2
\\ & \qquad
\,+\, 0.150\, \UL{\sigma}_{\textsf x}^1\UL{\sigma}_{\textsf z}^2 \,-\,
0.039\, \UL{\sigma}_{\textsf y}^1\UL{\sigma}_{\textsf z}^2
\,+\, 0.111\, \UL{\sigma}_{\textsf z}^1\UL{\sigma}_{\textsf x}^2 \,+\,
0.106\, \UL{\sigma}_{\textsf z}^1\UL{\sigma}_{\textsf y}^2
\end{aligned}
\notag \\ \UL L_3 ~\approx~ &
+0.046\, \UL{\sigma}_{\textsf z}^1 \,-\, 0.026\, \UL{\sigma}_{\textsf z}^2 \,-\,
0.057\, \UL{\sigma}_{\textsf x}^1\UL{\sigma}_{\textsf y}^2
\,-\, 0.266\, \UL{\sigma}_{\textsf z}^1\UL{\sigma}_{\textsf z}^2
\\ \notag \UL L_4 ~\approx~ &
\begin{aligned}[t] &
-0.024\, \UL{\sigma}_{\textsf x}^1 \,-\, 0.006\, \UL{\sigma}_{\textsf y}^2
\,-\, 0.081\, \UL{\sigma}_{\textsf x}^2 \,-\, 0.077\, \UL{\sigma}_{\textsf y}^2
\\ & \qquad
\,-\, 0.155\, \UL{\sigma}_{\textsf x}^1\UL{\sigma}_{\textsf z}^2 \,-\,
0.193\, \UL{\sigma}_{\textsf y}^1\UL{\sigma}_{\textsf z}^2
\,+\, 0.002\, \UL{\sigma}_{\textsf z}^1\UL{\sigma}_{\textsf x}^2 \,-\,
0.012\, \UL{\sigma}_{\textsf z}^1\UL{\sigma}_{\textsf y}^2
\end{aligned}
\\ \notag \UL L_5 ~\approx~ &
\begin{aligned}[t] &
-0.017\, \UL{\sigma}_{\textsf x}^1 \,-\, 0.060\, \UL{\sigma}_{\textsf y}^2
\,-\, 0.071\, \UL{\sigma}_{\textsf x}^2 \,-\, 0.090\, \UL{\sigma}_{\textsf y}^2
\\ & \qquad
\,+\, 0.090\, \UL{\sigma}_{\textsf x}^1\UL{\sigma}_{\textsf z}^2 \,-\,
0.001\, \UL{\sigma}_{\textsf y}^1\UL{\sigma}_{\textsf z}^2
\,-\, 0.183\, \UL{\sigma}_{\textsf z}^1\UL{\sigma}_{\textsf x}^2 \,-\,
0.118\, \UL{\sigma}_{\textsf z}^1\UL{\sigma}_{\textsf y}^2
\end{aligned}
\end{align}
It can be shown that $L_3$ contributes about $0.15\textrm s^{-1}$
to the decay rates of the single-quantum coherences (single-spin flips),
bringing down the decay time $T_2 \approx 2.6$s and, save for some
small cross terms among in the single quantum block, rather little else.

The superoperators corresponding to each of the remaining
$9$ Lindblads separately all contained significant cross terms between
the populations and the zero or double quantum coherences,
in violation of the secular approximation \cite{Ernst}.
Only on summing over all of them do these nonphysical cross-terms cancel out,
leaving a largely diagonal relaxation superoperator behind:
the ratio of the mean-square value of the off-diagonal
entries of $\mathcal R$ to that of the diagonal entries was
$1.3$\% in the transition basis and $3.8$\% in the Zeeman;
the latter dropped to $1.8$\% on excluding the block corresponding
to $T_1$ relaxation of the populations (\emph{vide infra}).
The nonphysical nature of most of the Lindblads is clearly
an artifact of the way that our procedure for calculating
them forces them to be orthogonal and minimal in number.
In order to physically interpret
the dominant relaxation processes, we therefore focus
our attention first on $T_1$ relaxation among the populations
(diagonal entries of the density matrix in the Zeeman basis),
along with the associated nonadiabatic $T_2$ relaxation,
and then try to account for the remaining $T_2$ relaxation
via simple adiabatic, albeit correlated, processes.

Based on Hadamard operator calculations that can be found in the appendix,
and by treating the $T_1$ and $T_2$ relaxation processes separetely,
we found that the four Lindblad operators which describe the $T_1$
relaxation of the first spin may be replaced by the Hermitian operators:
\begin{align}
L_{T_1}^{\textsf{x1}} ~=~ & 
\sqrt{0.1532}\, \tfrac12\, \sigma_{\textsf x}^1 \notag \\
L_{T_1}^{\textsf{y1}} ~=~ &
\sqrt{0.1532}\, \tfrac12\, \sigma_{\textsf y}^1 \\ \notag
L_{T_1}^{\textsf{xz}} ~=~ & 
\sqrt{0.1532}\, \tfrac12\, \sigma_{\textsf x}^1
\sigma_{\textsf z}^2 \\ \notag
L_{T_1}^{\textsf{yz}} ~=~ & 
\sqrt{0.1532}\, \tfrac12\, \sigma_{\textsf y}^1
\sigma_{\textsf z}^2
\end{align}
and for the second spin:
\begin{equation} \begin{split}
L_{T_1}^{\textsf{x2}} ~=~ \sqrt{0.1528}\, \tfrac12\, \sigma_{\textsf x}^2 ,\quad &
L_{T_1}^{\textsf{y2}} ~=~ \sqrt{0.1528}\, \tfrac12\, \sigma_{\textsf y}^2 , \\
L_{T_1}^{\textsf{zx}} ~=~ \sqrt{0.1528}\, \tfrac12\, \sigma_{\textsf z}^1
\sigma_{\textsf x}^2 ,\quad &
L_{T_1}^{\textsf{zy}} ~=~ \sqrt{0.1528}\, \tfrac12\, \sigma_{\textsf z}^1
\sigma_{\textsf y}^2 ~.
\end{split} \end{equation}
In addition, the near-degeneracy of the $(1,4)$ and $(2,3)$ rates
in the relaxation superoperator in the Zeeman basis 
can be used to combine the associated Lindblad operators into four
multiple-quantum $T_1$ Lindblad operators based on the average rate:
\begin{equation} \begin{split}
L_{T_1}^{\textsf{xx}} ~=~ \sqrt{0.0252}\, \tfrac12 \, \sigma_{\textsf{x}}^1 \sigma_{\textsf{x}}^2 ,\quad &
L_{T_1}^{\textsf{xy}} ~=~ \sqrt{0.0252}\, \tfrac12 \, \sigma_{\textsf{x}}^1\sigma_{\textsf{y}}^2 , \\
L_{T_1}^{\textsf{yx}} ~=~ \sqrt{0.0252}\, \tfrac12 \, \sigma_{\textsf{y}}^1\sigma_{\textsf{x}}^2 ,\quad &
L_{T_1}^{\textsf{yy}} ~=~ \sqrt{0.0252}\, \tfrac12 \, \sigma_{\textsf{y}}^1\sigma_{\textsf{y}}^2 ~.
\end{split} \end{equation}

By working through some examples, it may be seen that the sum of
the Lindbladian superoperators for each of the three sets of four
Lindblad operators above also causes all the \emph{off}-diagonal
entries of $\UL{\smash\rho}_\Delta^{\textsf{zee}}$
to decay with the same rate constant $1/T_1$.
This corresponds to \emph{nonadiabatic} $T_2$ relaxation.
Likewise, based on the results above, we substracted the nonadiabatic
$T_2$ contribution above to get the adiabatic $T_2$ contribution and thereby obtained
the following three Lindblad operators:
\begin{equation} \begin{array}{c}
L_{\,T_2}^{\,\textsf{ad1}} ~=~ \sqrt{0.9560}\; \tfrac1{\sqrt8}\;
\big( \sigma_{\,\mathsf z}^1 + \sigma_{\,\mathsf z}^2 \big) ,\quad
L_{\,T_2}^{\,\textsf{ad2}} ~=~ \sqrt{0.1721}\; \tfrac1{\sqrt8}\;
\big( \sigma_{\,\mathsf z}^1 - \sigma_{\,\mathsf z}^2 \big) ,
\\[6pt]
L_{\,T_2}^{\,\textsf{ad3}} ~=~ \sqrt{0.2931}\; \tfrac12\;
\sigma_{\,\mathsf z}^1\, \sigma_{\,\mathsf z}^2 ~.
\end{array} \end{equation}
These correspond to totally correlated, totally anticorrelated,
and pure single-quantum $T_2$ relaxation, respectively \cite{HavHad}.
As a result, the Hadamard product formalism allowed us to obtain a
description with a clearer physical interpretation. However, this left a slight
discrepancy between the new results and the original data, because
some of the assumptions made, such as the near degeneracy of some
decay rates mentioned above, were only approximations.
In order to assess how much of the total experimental superoperator
$\mathcal R$ was accounted for by our simplifications, 
we computed $\UL R_{\,T_1}^{\,\textsf{zee}}$ and
$\UL R_{\,T_2}^{\,\textsf{zee}}$ from the above
Lindblad matrices, obtained as described in the Appendix,
and computed the relative discrepancy in the superoperators,
\begin{equation}
\frac{\| \UL{\mathcal R}^{\textsf{zee}} - \UL{\smash{\mathrm{Diag}}}
(\mket{\UL R_{\,T_2}^{\,\textsf{zee}}}) - \UL{\smash{\mathrm{Choi}}}
(\mket{\UL R_{\,T_1}^{\,\textsf{zee}}}) \|^2}
{\| \UL{\mathcal R}^{\,\textsf{zee}} \|^2} ~,
\end{equation}
where "zee" indicates that the above quantities were expressed in the 
Zeeman basis, and where $\UL{\mathrm{Choi}}$ is a function which distributes
the entries of its $4\times4$ argument over the corresponding entries
of a $16\times16$ matrix and sets all the other entries to zero.
This gave a value of $6.3$\%, indicating that the approximations
introduced in the Appendix were sufficient to account for
a large majority of the observed relaxation dynamics.

\section{Conclusion}
In this paper, we have demonstrated a robust procedure by which one can derive a 
set of Lindblad operators which collectively account for a Markovian quantum process,
without any prior assumptions regarding the nature of the process beyond the physical
necessity of complete positivity. This procedure should be widely useful in studies
of dissipative quantum processes. In the appendix, we have further shown how one can use 
one's physical insight into the particular system in question to derive the physical
"noise generators" of the system. We believe this two-step process is illustrative
of how Quantum Process Tomography on many physically distinct kinds of quantum processes
should be done.

\bigskip\centerline{\textit{Acknowledgments}}
This work was supported by ARO through grants DAAD19-01-1-0519, DAAD19-01-1-0678,
DARPA MDA972-01-1-0003 and NSF EEC-0085557. We thank L. Viola, E. M. Fortunato and J.
Emerson for valuable discussions. Correspondence and requests for materials should be
addressed to T. F.Havel (email: \texttt{tfhavel@mit.edu}).

\section{Appendix}
In this appendix, we shall derive a more compact and direct representation of the $T_2$
relaxation processes operative in 2,3-dibromothiophene via the so-called ``Hadamard relaxation matrix''
\cite{HavHad} ({\it vide infra}).

The populations block of the relaxation superoperator
corresponds to indices $1$ through $4$ in the transition basis (see Table I.) and
the nonzero entries of $\mket{\UL I}\mbra{\UL I}$ in the Zeeman.
The values obtained from the completely positive least-squares
fit shown in FIG.~\ref{fig:results} are
\begin{equation}
\begin{aligned}[b] &
\begin{matrix} \hspace{1.5em} \textsf{iden} & \hspace{1em}
\UL{\sigma}_{\textsf{z}}^1 & \hspace{1.2em} \UL{\sigma}_{\textsf{z}}^2 &
\hspace{0.7em} \UL{\sigma}_{\textsf{z}}^1\UL{\sigma}_{\textsf{z}}^2
\end{matrix} \\
\UL R_{\,T_1}^{\,\textsf{tra}} ~=~ &
\left[ \begin{smallmatrix}
~0.0000 &~0.0000 &~0.0000 &~0.0000 \\[3pt]
~0.0000 &~0.1780 &-0.0002 &~0.0089 \\[3pt]
~0.0000 &-0.0002 &~0.1784 &-0.0097 \\[3pt]
~0.0000 &~0.0089 &-0.0097 &~0.3061
\end{smallmatrix} \right]
\end{aligned}
~\leftrightarrow~
\begin{aligned}[b] &
\begin{matrix} \hspace{1.5em} \mket{\uparrow\uparrow} & \hspace{0.75em}
\mket{\uparrow\downarrow} & \hspace{0.75em} \mket{\downarrow\uparrow} &
\hspace{0.75em} \mket{\downarrow\downarrow}
\end{matrix} \\ &
\left[ \begin{smallmatrix}
~0.3301 &-0.1435 &-0.1617 &-0.0249 \\[3pt]
-0.1435 &~0.3129 &-0.0254 &-0.1440 \\[3pt]
-0.1617 &-0.0254 &~0.3500 &-0.1630 \\[3pt]
-0.0249 &-0.1440 &-0.1629 &~0.3319
\end{smallmatrix} \right]
~=~ \UL R_{\,T_1}^{\,\textsf{zee}}
\end{aligned}
\end{equation}
in the transition (left) as well as the Zeeman (right) bases.
The matrices $\UL R_{\,T_1}^{\,\textsf{tra}}$ and $\UL R_{\,T_1}^{\,\textsf{zee}}$
are related by the \emph{Hadamard transform} $\UL W$ \cite{Tseng,HavHad},
\begin{equation}
\UL W ~\equiv~ \tfrac12 \left[ \begin{smallmatrix} \rule[1.5ex]{0pt}{0pt}
~1&~1&~1&~1\\[3pt] ~1&~1&-1&-1\\[3pt] ~1&-1&~1&-1\\[3pt] ~1&-1&-1&~1
\rule[-1ex]{0pt}{0pt}\end{smallmatrix} \right]
\end{equation}
that is
\begin{equation}
2\, \UL W\, \UL R_{\,T_1}^{\,\textsf{zee}}\, \UL W ~=~
\UL R_{\,T_1}^{\,\textsf{tra}} \quad\leftrightarrow\quad
\UL R_{\,T_1}^{\,\textsf{zee}} ~=~ \tfrac12\, \UL W\,
\UL R_{\,T_1}^{\,\textsf{tra}}\, \UL W ~,
\end{equation}
since $\UL W^2 = \UL I$. In the absence of cross-correlation,
symmetry considerations imply that $\UL R_{\,T_1}^{\,\textsf{zee}}$
should be centrosymmetric \cite{Solomon,Freeman}, and hence we shall use
the symmetrized version $\frac12\big( \UL R_{\,T_1}^{\,\textsf{zee}}
\,+\, \UL{\sigma}_{\textsf x}^1\UL{\sigma}_{\textsf x}^2\,
\UL R_{\,T_1}^{\,\textsf{zee}}\,
\UL{\sigma}_{\textsf x}^1\UL{\sigma}_{\textsf x}^2 \big)$
and its Hadamard transform in the following, which are
\begin{equation}
\UL R_{\,T_1}^{\,\textsf{tra}} ~=~
\left[ \begin{smallmatrix}
 0.0000 & -0.0000 & -0.0000 &  0.0000 \\[3pt]
 0.0000 &  0.1780 & -0.0002 &  0.0000 \\[3pt]
 0.0000 & -0.0002 &  0.1784 &  0.0000 \\[3pt]
 0.0000 &  0.0000 &  0.0000 &  0.3061
\end{smallmatrix} \right]
~\leftrightarrow~
\left[ \begin{smallmatrix}
~0.3310 & -0.1532 & -0.1528 & -0.0249 \\[3pt]
-0.1532 & ~0.3315 & -0.0254 & -0.1528 \\[3pt]
-0.1528 & -0.0254 & ~0.3315 & -0.1532 \\[3pt]
-0.0249 & -0.1528 & -0.1532 & ~0.3310
\end{smallmatrix} \right]
~=~ \UL R_{\,T_1}^{\,\textsf{zee}} ~.
\end{equation}
As noted in the main paper, the $T_1$ of both spins is $\sim\!5.6\,$s, while the
NOE rate (connecting $\sigma_{\textsf{z}}^1$ and $\sigma_{\textsf{z}}^2$
in the transition basis) is negligibly small ($-0.0002$).

The entries of $-\UL R_{\,T_1}^{\,\textsf{zee}}$ are equal to the diagonal
entries of the Choi matrix $\UL{\mathcal C}$ of $-\UL{\mathcal R}^{\textsf{zee}}$,
and the \emph{off}-diagonal entries of $\UL R_{\,T_1}^{\,\textsf{zee}}$ are
the \emph{only} non-negligible entries in their respective rows and columns of $\UL{\mathcal C}$.
Therefore, they are eigenvalues of the Choi matrix as well as its
projection $\UL{\mathcal E}\, \UL{\mathcal C}\, \UL{\mathcal E}$,
and their corresponding eigenvectors are elementary unit
vectors $\mket{k}\mket{j}$ relative to the Zeeman basis.
It follows that the Lindblad operator for the $(j,k)$-th off-diagonal entry
of $\UL R_{\,T_1}^{\,\textsf{zee}}$ may be written as $\UL L_{T_1}^{jk}
\equiv (-\mbra{j}\, \UL R_{\,T_1}^{\,\textsf{zee}}\, \mket{k})^{1/2}\,
\mket{k} \mbra{j}$, and its contribution to $\UL{\smash{\dot\rho}}_\Delta^{\,\textsf{zee}}$
is given by
\begin{equation}
L_{T_1}^{jk}(\UL{\smash{\dot\rho}}_\Delta^{\,\textsf{zee}}) ~\equiv~
-\mbra{j}\, \UL R_{\,T_1}^{\,\textsf{zee}}\, \mket{k}\, \Big(
\mket{k}\mbra{j}\, \UL{\smash{\rho}}_{\,\Delta}^{\,\textsf{zee}}\, \mket{j}\mbra{k}
\,-\, \tfrac12\, \mket{j}\mbra{j}\, \UL{\smash{\rho}}_{\,\Delta}^{\,\textsf{zee}}
\,-\, \tfrac12\, \UL{\smash{\rho}}_{\,\Delta}^{\,\textsf{zee}}\, \mket{j}\mbra{j} \Big) ~.
\end{equation}
The symmetry of $\UL R_{\,T_1}^{\,\textsf{zee}}$ implies that
the eigenvalues of $\UL{\mathcal C}$ corresponding to single spin
flips (the so-called single-quantum coherences) are four-fold
degenerate, and hence we can replace their elementary unit
eigenvectors by arbitrary unitary linear combinations thereof.
For example, the four Lindblad operators which describe the $T_1$
relaxation of the first spin may be replaced by the Hermitian operators:
\begin{align}
L_{T_1}^{\textsf{x1}} ~=~ & \sqrt{0.1532}\, \tfrac12\, \big(
\mket2\mbra0 + \mket0\mbra2 + \mket3\mbra1 + \mket1\mbra3 \big)
~=~ \sqrt{0.1532}\, \tfrac12\, \UL{\sigma}_{\textsf x}^1 \notag \\
L_{T_1}^{\textsf{y1}} ~=~ & \sqrt{0.1532}\, \tfrac{\displaystyle\imath}2\, \big(
\mket2\mbra0 - \mket0\mbra2 + \mket3\mbra1 - \mket1\mbra3 \big)
~=~ \sqrt{0.1532}\, \tfrac12\, \UL{\sigma}_{\textsf y}^1 \\ \notag
L_{T_1}^{\textsf{xz}} ~=~ & \sqrt{0.1532}\, \tfrac12\, \big(
\mket2\mbra0 + \mket0\mbra2 - \mket3\mbra1 - \mket1\mbra3 \big)
~=~ \sqrt{0.1532}\, \tfrac12\, \UL{\sigma}_{\textsf x}^1
\UL{\sigma}_{\textsf z}^2 \\ \notag
L_{T_1}^{\textsf{yz}} ~=~ & \sqrt{0.1532}\, \tfrac{\displaystyle\imath}2\, \big(
\mket2\mbra0 - \mket0\mbra2 - \mket3\mbra1 + \mket1\mbra3 \big)
~=~ \sqrt{0.1532}\, \tfrac12\, \UL{\sigma}_{\textsf y}^1
\UL{\sigma}_{\textsf z}^2
\end{align}
In a similar fashion, we may take those describing
$T_1$ relaxation of the second spin to be
\begin{equation} \begin{split}
L_{T_1}^{\textsf{1x}} ~=~ \sqrt{0.1528}\, \tfrac12\, \UL{\sigma}_{\textsf x}^2 ,\quad &
L_{T_1}^{\textsf{1y}} ~=~ \sqrt{0.1528}\, \tfrac12\, \UL{\sigma}_{\textsf y}^2 , \\
L_{T_1}^{\textsf{zx}} ~=~ \sqrt{0.1528}\, \tfrac12\, \UL{\sigma}_{\textsf z}^1
\UL{\sigma}_{\textsf x}^2 ,\quad &
L_{T_1}^{\textsf{zy}} ~=~ \sqrt{0.1528}\, \tfrac12\, \UL{\sigma}_{\textsf z}^1
\UL{\sigma}_{\textsf y}^2 ~.
\end{split} \end{equation}
In addition, the near-degeneracy of the $(1,4)$ and $(2,3)$ rates
can be used to combine the associated Lindblad operators into four
multiple-quantum $T_1$ Lindblad operators based on the average rate:
\begin{equation} \begin{split}
\UL L_{T_1}^{\textsf{xx}} ~=~ \sqrt{0.0252}\, \tfrac12 \, \UL{\sigma}_{\textsf{x}}^1\UL{\sigma}_{\textsf{x}}^2 ,\quad &
\UL L_{T_1}^{\textsf{xy}} ~=~ \sqrt{0.0252}\, \tfrac12 \, \UL{\sigma}_{\textsf{x}}^1\UL{\sigma}_{\textsf{y}}^2 , \\
\UL L_{T_1}^{\textsf{yx}} ~=~ \sqrt{0.0252}\, \tfrac12 \, \UL{\sigma}_{\textsf{y}}^1\UL{\sigma}_{\textsf{x}}^2 ,\quad &
\UL L_{T_1}^{\textsf{yy}} ~=~ \sqrt{0.0252}\, \tfrac12 \, \UL{\sigma}_{\textsf{y}}^1\UL{\sigma}_{\textsf{y}}^2 ~.
\end{split} \end{equation}

By working through some examples, it may be seen that the sum of
the Lindbladian superoperators for each of the three sets of four
Lindblad operators above also causes all the \emph{off}-diagonal
entries of $\UL{\smash\rho}_\Delta^{\textsf{zee}}$
to decay with the same rate constant $1/T_1$.
This corresponds to \emph{nonadiabatic} $T_2$ relaxation.
Because $\UL R_{\,T_1}^{\,\textsf{zee}}$ does not act on its
off-diagonal entries, it may also be seen that if one takes the Lindblads 
$(\mbra{k}\UL R_{\,T_1}^{\,\textsf{zee}}\mket{k})^{1/2}\,\mket{k}\!\mbra{k}$
of the four diagonal entries of $\UL R_{\,T_1}^{\,\textsf{zee}}$ and
\emph{subtracts} their superoperators from those for the off-diagonal,
this must exactly cancel the nonadiabatic $T_2$ decay.
Formally, however, the negative of a Lindbladian
superoperator is not a Lindbladian superoperator,
and in any case we do not really want to cancel
the nonadiabatic $T_2$, since it actually occurs.
In order to avoid accounting for the nonadiabatic $T_2$ twice,
it is nevertheless necessary to write down a set of Lindblad
operators for it alone, without any $T_1$ relaxation.
Once again, on using the near equality of the diagonal
entries of $\UL R_{\,T_1}^{\,\textsf{zee}}$ to replace
them by their average and taking suitable unitary linear
combinations of the diagonal Lindblads $\mket{j}\mbra{j}$,
we obtain
\begin{equation} \begin{split}
\UL L_{T_1}^{\textsf{na0}} ~=~ \sqrt{0.3312}\, \tfrac12\, \UL I ,\quad &
\UL L_{T_1}^{\textsf{na1}} ~=~ \sqrt{0.3312}\, \tfrac12\, \UL{\sigma}_{\textsf{z}}^1
,~ \\
\UL L_{T_1}^{\textsf{na2}} ~=~ \sqrt{0.3312}\, \tfrac12\, \UL{\sigma}_{\textsf{z}}^2
,\quad & \UL L_{T_1}^{\textsf{na3}} ~=~ \sqrt{0.3312}\, \tfrac12\,
\UL{\sigma}_{\textsf{z}}^1\UL{\sigma}_{\textsf{z}}^2 ~. \label{eq:nat2}
\end{split} \end{equation}
The Lindbladian superoperator of the first of these is obviously
$L_ {T_1}^{\textsf{na0}}(\UL{\smash\rho}_\Delta) = \UL 0$,
and so need not be considered further.

We now turn our attention to the diagonal entries of the $16\times16$
Zeeman relaxation superoperator $\UL{\mathcal R}^{\textsf{zee}}$,
which we shall arrange in a $4\times4$ matrix of relaxation
rate constants of the corresponding entries of the density
matrix $\UL{\smash{\rho}}_{\,\Delta}^{\,\textsf{zee}}$.
It can be shown that this $4\times4$ matrix
\begin{equation}
\UL R_{\textsf{diag}}^{\,\textsf{zee}} ~\equiv~ \begin{bmatrix}
\mbra{00}\UL{\mathcal R}^{\textsf{zee}}\mket{00} &
\mbra{01}\UL{\mathcal R}^{\textsf{zee}}\mket{01} &
\mbra{02}\UL{\mathcal R}^{\textsf{zee}}\mket{02} &
\mbra{03}\UL{\mathcal R}^{\textsf{zee}}\mket{03} \\
\mbra{10}\UL{\mathcal R}^{\textsf{zee}}\mket{10} &
\mbra{11}\UL{\mathcal R}^{\textsf{zee}}\mket{11} &
\mbra{12}\UL{\mathcal R}^{\textsf{zee}}\mket{12} &
\mbra{13}\UL{\mathcal R}^{\textsf{zee}}\mket{13} \\
\mbra{20}\UL{\mathcal R}^{\textsf{zee}}\mket{20} &
\mbra{21}\UL{\mathcal R}^{\textsf{zee}}\mket{21} &
\mbra{22}\UL{\mathcal R}^{\textsf{zee}}\mket{22} &
\mbra{23}\UL{\mathcal R}^{\textsf{zee}}\mket{23} \\
\mbra{30}\UL{\mathcal R}^{\textsf{zee}}\mket{30} &
\mbra{31}\UL{\mathcal R}^{\textsf{zee}}\mket{31} &
\mbra{32}\UL{\mathcal R}^{\textsf{zee}}\mket{32} &
\mbra{33}\UL{\mathcal R}^{\textsf{zee}}\mket{33} \end{bmatrix}
\label{eq:rt1}
\end{equation}
is also a  symmetric submatrix of the Choi matrix $\UL{\mathcal C}$
associated with $\UL{\mathcal R}^{\textsf{zee}}$ (up to sign), specifically
\begin{equation}
-\UL R_{\textsf{diag}}^{\,\textsf{zee}} ~=~ \begin{bmatrix}
\mbra{00}\,\UL{\mathcal C}\,\mket{00} & \mbra{00}\,\UL{\mathcal C}\,\mket{11} &
\mbra{00}\,\UL{\mathcal C}\,\mket{22} & \mbra{00}\,\UL{\mathcal C}\,\mket{33} \\
\mbra{11}\,\UL{\mathcal C}\,\mket{00} & \mbra{11}\,\UL{\mathcal C}\,\mket{11} &
\mbra{11}\,\UL{\mathcal C}\,\mket{22} & \mbra{11}\,\UL{\mathcal C}\,\mket{33} \\
\mbra{22}\,\UL{\mathcal C}\,\mket{00} & \mbra{22}\,\UL{\mathcal C}\,\mket{11} &
\mbra{22}\,\UL{\mathcal C}\,\mket{22} & \mbra{22}\,\UL{\mathcal C}\,\mket{33} \\
\mbra{33}\,\UL{\mathcal C}\,\mket{00} & \mbra{33}\,\UL{\mathcal C}\,\mket{11} &
\mbra{33}\,\UL{\mathcal C}\,\mket{22} & \mbra{33}\,\UL{\mathcal C}\,\mket{33}
\end{bmatrix} ~.
\end{equation}
The diagonal of $\UL R_{\textsf{diag}}^{\,\textsf{zee}}$ is thus
the same as the diagonal of $\UL R_{\,T_1}^{\,\textsf{zee}}$.
Since we have already found a set of Lindblad operators which
fully account for the effects of $\UL R_{\,T_1}^{\,\textsf{zee}}$
on $\UL{\smash\rho}_{\,\Delta}^{\,\textsf{zee}}$,
we will now focus our attention on its off-diagonal entries
by defining a new matrix $\UL R_{\,T_2}^{\textsf{zee}}$ which
is the same as $\UL R_{\textsf{diag}}^{\,\textsf{zee}}$ save for
its four diagonal entreis, which are set to zero as indicated below:
\begin{equation}
\UL R_{\,T_2}^{\textsf{zee}} ~\equiv~ \UL R_{\textsf{diag}}^{\,\textsf{zee}}
~-~ \UL{\smash{\mathrm{Diag}}}\big(\, \mbra{kk}\UL{\mathcal R}^{\textsf{zee}}\mket{kk}
\,\big|\, k=0,\ldots,3\, \big) .
\end{equation}

As implied by our notation, $\UL R_{\,T_2}^{\,\textsf{zee}}$
contains all the information regarding $T_2$ relaxation
processes that is contained in our full, but \emph{diagonal},
relaxation superoperator $\UL{\mathcal R}^{\textsf{zee}}$, and
in a considerably more compact and easily understood form.
Unlike $\UL R_{\,T_1}^{\,\textsf{zee}}$,
which acts on the column vector $\UL{\smash{\mathrm{diag}}}
(\UL{\smash\rho}_{\,\Delta}^{\,\textsf{zee}})$ of diagonal entries
by matrix multiplication, $\UL R_{\,T_2}^{\,\textsf{zee}}$ acts on
$\UL{\smash\rho}_{\,\Delta}^{\,\textsf{zee}}$ simply by taking the
products of all corresponding pairs of entries, one from each matrix,
just as these entries are multiplied together in the full
matrix-vector product $\UL{\mathcal R}^{\textsf{zee}}\,
\mket{\UL{\smash\rho}_\Delta^{\textsf{zee}}}$.
This ``entrywise'' matrix multiplication, commonly
known as the \emph{Hadamard product} \cite{HornJohn},
has already been shown to be a powerful means of describing
``simple'' $T_2$ relaxation processes \cite{HavHad}
(that is, processes not involving cross-relaxation).
The Hadamard product will be denoted in the following as:
\begin{equation}
\UL{\smash{\dot\rho}}_{\,\Delta}^{\,\textsf{zee}} ~=~ -\UL R_{\,T_2}^{\,\textsf{zee}}
\odot \UL{\smash\rho}_{\,\Delta}^{\,\textsf{zee}} ~\equiv~ -\big[\, \mbra{j}\, \rho_{\,\Delta}^{\,\textsf{zee}}\, \mket{k}\, \mbra{j}\, \UL R_{\,T_2}^{\,\textsf{zee}}\,
\mket{k}\, \big]_{j,k=0}^3 ~.
\end{equation}

Another important property of the matrix $\UL R_{\,T_2}^{\,\textsf{zee}}$
is that, since the overall projected Choi matrix $\UL{\mathcal E}\,
\UL{\mathcal C}\, \UL{\mathcal E}$ must be positive semidefinite,
the same is true of the projection of $-\UL R_{\,T_2}^{\,\textsf{zee}}$,
and the $4\times4$ block of the $16\times16$ projection matrix
$\UL{\mathcal E}$ that acts on $-\UL R_{\,T_2}^{\,\textsf{zee}}$ is
\begin{equation}
\UL E ~\equiv~ \UL I \,-\, \tfrac14\, \UL 1\, \UL 1^\top
~=~ \tfrac14\! \left[ \begin{smallmatrix} \rule[1.75ex]{0pt}{0pt} 
~3 & -1 & -1 & -1\\[3pt] -1 & ~3 & -1 & -1\\[3pt]
-1 & -1 & ~3 & -1\\[3pt] -1 & -1 & -1 & ~3
\rule[-0.75ex]{0pt}{0pt} \end{smallmatrix} \right] ~,
\end{equation}
where $\UL 1$ denotes a column vector of four $1$'s.
Moreover, the Lindblad operators for $T_2$ relaxation may be extracted
directly from $-\UL E\, \UL R_{\,T_2}^{\,\textsf{zee}}\, \UL E$ without
direct reference to the full superoperator's projected Choi matrix.

To see how this can be done, we first observe that given any two diagonal
matrices $\UL A$ \& $\UL C$ (assumed here to equal in dimension) with
column vectors of diagonal entries $\UL a = \UL{\smash{\mathrm{diag}}}(\UL A)$
\& $\UL c = \UL{\smash{\mathrm{diag}}}(\UL C)$, respectively, we have
\begin{equation}
\UL A\, \UL B\, \UL C ~=~ (\UL A\, \UL 1\, \UL 1^\top \UL C) \odot \UL B
~=~ (\UL a\, \UL c^\top) \odot \UL B
\end{equation}
for any other (not necessarily diagonal) matrix $\UL B$ of equal dimension.
It follows that the action on a density operator
$\rho$ of any Lindblad operator $L$, with respect to
a basis wherein its matrix $\UL L$ is real and diagonal,
can be expressed in terms of Hadamard products as
\begin{equation}
\UL{\smash{L(\rho)}} ~\equiv~ \UL L\, \UL{\smash\rho}\, \UL L \,-\, \tfrac12\,
\UL L^2 \UL{\smash\rho} \,-\, \tfrac12\, \UL{\smash\rho}\, \UL L^2
~=~ \big( \UL L\, \UL 1\, \UL 1^\top \UL L \,-\, \tfrac12\, \UL L^2\,
\UL 1\, \UL 1^\top -\, \tfrac12\, \UL 1\, \UL 1^\top \UL L^2 \big)
\odot\, \UL{\smash\rho} ~\equiv\, -\UL R_L \odot\, \UL{\smash\rho} ~,
\end{equation}
where $\UL R_L$ is called a \emph{Hadamard relaxation matrix} for $L$.
If multiple Lindblad operators act simultaneously,
the net Hadamard relaxation matrix is of course the
sum of those associated with the individual Lindblads.

Next, let us use the Lindblad operators for nonadiabatic $T_2$ relaxation
given in Eq.~(\ref{eq:nat2}) above to illustrate how we can go the other way,
that is derive these Lindblads from the corresponding Hadamard relaxation matrix.
If we let $\UL\ell_{\,T_2}^{\,\textsf{na}i} \equiv\UL{\smash{\mathrm{diag}}}
( \UL L_{T_2}^{\textsf{na}i} )$ be the column vectors formed from
the real diagonal entries of these Lindblad matrices and observe
that their Hadamard squares $\UL\ell_{\,T_2}^{\,\textsf{na}i} \odot
\UL\ell_{\,T_2}^{\,\textsf{na}i} = 0.3312\; \tfrac14\, \UL 1$ for $i=1,2,3$,
then this may be written as
\begin{align}
-\UL R_{\,T_2}^{\,\text{na}} ~\equiv~ & \begin{aligned}[t] &
\UL\ell_{T_2}^{\,\textsf{na1}}\, \bigl( \UL \ell_{T_2}^{\,\textsf{na1}} \bigr)^{\!\top} +~
\UL\ell_{T_2}^{\,\textsf{na2}}\, \bigl( \UL \ell_{T_2}^{\,\textsf{na1}} \bigr)^{\!\top} +~
\UL\ell_{T_2}^{\,\textsf{na1}}\, \bigl( \UL \ell_{T_2}^{\,\textsf{na1}} \bigr)^{\!\top}
\\ & -~ \tfrac12\, \bigl(
\UL\ell_{T_2}^{\,\textsf{na1}}\odot\UL \ell_{T_2}^{\,\textsf{na1}}
~+~ \UL\ell_{T_2}^{\,\textsf{na2}}\odot\UL \ell_{T_2}^{\,\textsf{na2}} ~+~
\UL\ell_{T_2}^{\,\textsf{na3}}\odot\UL \ell_{T_2}^{\,\textsf{na3}} \bigr)\, \UL 1^\top
\\ & -~ \tfrac12\; \UL 1\, \bigl(
\UL\ell_{T_2}^{\,\textsf{na1}}\odot\UL \ell_{T_2}^{\,\textsf{na1}}
~+~ \UL\ell_{T_2}^{\,\textsf{na2}}\odot\UL \ell_{T_2}^{\,\textsf{na2}} ~+~
\UL\ell_{T_2}^{\,\textsf{na3}}\odot\UL \ell_{T_2}^{\,\textsf{na3}} \bigr)^{\!\top}
\end{aligned}
\notag \\[3pt] =~ & \begin{aligned}[t] & \frac{0.3312}4 \Bigg(
\left[ \begin{smallmatrix} ~1\\[2pt] ~1\\[2pt] -1\\[2pt] -1 \end{smallmatrix} \right]
\left[ \begin{smallmatrix} \,1 & ~1 & -1 & -1 \end{smallmatrix} \right] ~+~
\left[ \begin{smallmatrix} ~1\\[2pt] -1\\[2pt] ~1\\[2pt] -1 \end{smallmatrix} \right]
\left[ \begin{smallmatrix} \,1 & -1 & ~1 & -1 \end{smallmatrix} \right] ~+~
\left[ \begin{smallmatrix} ~1\\[2pt] -1\\[2pt] -1\\[2pt] ~1 \end{smallmatrix} \right]
\left[ \begin{smallmatrix} \,1 & -1 & -1 & ~1 \end{smallmatrix} \right] \Bigg)
\\ & -~ \frac{0.3312}8 \Bigg(
\left[ \begin{smallmatrix} \rule[0.5ex]{0pt}{0pt} 1\\[2pt] 1\\[2pt] 1\\[2pt] 1
\rule[-0.5ex]{0pt}{0pt} \end{smallmatrix} \right] +
\left[ \begin{smallmatrix} \rule[0.5ex]{0pt}{0pt} 1\\[2pt] 1\\[2pt] 1\\[2pt] 1
\rule[-0.5ex]{0pt}{0pt} \end{smallmatrix} \right] +
\left[ \begin{smallmatrix} \rule[0.5ex]{0pt}{0pt} 1\\[2pt] 1\\[2pt] 1\\[2pt] 1
\rule[-0.5ex]{0pt}{0pt} \end{smallmatrix} \right]
\Bigg) \left[ \begin{smallmatrix} \,1 & \,1 & \,1 & \,1 \end{smallmatrix} \right]
~-~ \frac{0.3312}8
\left[ \begin{smallmatrix} \rule[0.5ex]{0pt}{0pt} 1\\[2pt] 1\\[2pt] 1\\[2pt] 1
\rule[-0.5ex]{0pt}{0pt} \end{smallmatrix} \right]
\begin{pmatrix}
\quad\, \left[ \begin{smallmatrix} \,1 & \,1 & \,1 & \,1 \end{smallmatrix} \right]
\\[-7pt] +~ \left[ \begin{smallmatrix} \,1 & \,1 & \,1 & \,1 \end{smallmatrix} \right]
\\[-7pt] +~ \left[ \begin{smallmatrix} \,1 & \,1 & \,1 & \,1 \end{smallmatrix} \right]
\end{pmatrix}
\end{aligned}
\\[3pt] \notag =~ & 0.3312 \big(\, \UL E ~-~ \tfrac34\, \UL 1\, \UL 1^\top \big)
~=\; -\,0.3312 \left[ \begin{smallmatrix}~0~&~1~&~1~&~1~\\[3pt]~1~&~0~&~1~&~1~\\[3pt]
~1~&~1~&~0~&~1~\\[3pt]~1~&~1~&~1~&~0~\end{smallmatrix} \right] ~.
\end{align}
Noting that $\UL E^2 = \UL E$ and that the row and column sums of $\UL E$ are zero,
it may be seen that the projection $-\UL E\, \UL R_{T_2}^{\,\textsf{na}}\, \UL E$
simply removes the last two terms involving the Hadamard squares from the above,
which are proportional to $\UL 1\,\UL 1^\top$,
leaving only $-\UL E\, \UL R_{T_2}^{\,\textsf{na}}\, \UL E = 0.3312\, \UL E$ behind.
Because the vectors $\UL\ell_{T_2}^{\,\textsf{na}i}$ are mutually orthogonal
and all their squared norms are $\|\UL\ell_{T_2}^{\,\textsf{na}i}\|^2 = 0.3312$,
upon normalization they actually become the eigenvectors of $\UL E$ that are
associated with its one nonzero, but triply degenerate, eigenvalue of unity.

From this we see that the nonzero entries of the diagonal Lindblad
matrices $\UL L_{\,T_2}^{\,\textsf{na}i}$ are the entries of the
eigenvectors of $-\UL E\, \UL R_{\,T_2}^{\,\textsf{na}}\, \UL E$
times the square roots of their respective eigenvalues,
much as general Lindblad matrices may be obtained from the
eigenvalues and eigenvectors of the projected Choi matrix
$\UL{\mathcal E}\, \UL{\mathcal C}\, \UL{\mathcal E}$.
The numerical values of the entries of $\UL R_{\,T_2}^{\,\textsf{zee}}$, as
extracted from the experimental superoperator in FIG.~\ref{fig:results}, are:
\begin{equation}
\UL R_{\,T_2}^{\,\textsf{zee}} ~=~
\left[ \begin{smallmatrix} \rule[1.7ex]{0pt}{0pt}
0 & 0.7890 & 0.7757 & 1.2872 \\[6pt]
0.7890 & 0 & 0.5033 & 0.7426 \\[6pt]
0.7757 & 0.5033 & 0 & 0.7283 \\[6pt]
1.2872 & 0.7426 & 0.7283 & 0
\rule[-1ex]{0pt}{0pt} \end{smallmatrix} \right]
~\textrm s^{-1} \,.
\end{equation}
It is easily seen that $\UL R_{\,T_2}^{\,\textsf{zee}}$,
like $\UL R_{\,T_1}^{\,\textsf{zee}}$, must be centrosymmetric,
and if we likewise symmetrize and subtract the above
nonadiabatic $T_2$ Hadamard relaxation matrix, we get
\begin{equation}
\UL R_{\,T_2}^{\,\textsf{ad}} ~\equiv~
\UL R_{\,T_2}^{\,\textsf{zee}} \,-\, \UL R_{\,T_2}^{\,\textsf{na}}
~=~ \left[ \begin{smallmatrix} \rule[1.7ex]{0pt}{0pt}
0 & 0.4274 & 0.4279 & 0.9560 \\[6pt]
0.4274 & 0 & 0.1721 & 0.4279 \\[6pt]
0.4279 & 0.1721 & 0 & 0.4274 \\[6pt]
0.9560 & 0.4279 & 0.4274 & 0
\rule[-1ex]{0pt}{0pt} \end{smallmatrix} \right] ~.
\end{equation}
The nonzero eigenvalues and associated eigenvectors of
$-\UL E\, \UL R_{\,T_2}^{\,\textsf{ad}}\, \UL E$ are
(to within $1$\%)
\begin{equation}
\Bigg( 0.9560 ,~ \tfrac1{\sqrt2}\! \left[ \begin{smallmatrix}
~1\\[3pt] ~0\\[3pt] ~0\\[3pt] -1 \end{smallmatrix} \right] \Bigg), \quad
\Bigg( 0.2913 ,~ \tfrac1{2}\! \left[ \begin{smallmatrix}
~1\\[3pt] \!-1\\[3pt] \!-1\\[3pt] ~1 \end{smallmatrix} \right] \Bigg), \quad
\Bigg( 0.1721 ,~ \tfrac1{\sqrt2}\! \left[ \begin{smallmatrix}
~0\\[3pt] ~1\\[3pt] \!-1\\[3pt] ~0 \end{smallmatrix} \right] \Bigg),
\end{equation}
which correspond to a system of three Lindblad operators
for the adiabatic $T_2$ relaxation, namely
\begin{equation} \begin{array}{c}
\UL L_{\,T_2}^{\,\textsf{ad1}} ~=~ \sqrt{0.9560}\; \tfrac1{\sqrt8}\;
\big( \UL{\sigma}_{\,\mathsf z}^1 + \UL{\sigma}_{\,\mathsf z}^2 \big) ,\quad
\UL L_{\,T_2}^{\,\textsf{ad2}} ~=~ \sqrt{0.1721}\; \tfrac1{\sqrt8}\;
\big( \UL{\sigma}_{\,\mathsf z}^1 - \UL{\sigma}_{\,\mathsf z}^2 \big) ,
\\[6pt]
\UL L_{\,T_2}^{\,\textsf{ad3}} ~=~ \sqrt{0.2931}\; \tfrac12\;
\UL{\sigma}_{\,\mathsf z}^1\, \UL{\sigma}_{\,\mathsf z}^2 ~.
\end{array} \end{equation}
These correspond to totally correlated, totally anticorrelated,
and pure single-quantum, i.e. dipolar $T_2$ relaxation, respectively \cite{HavHad}.

\end{document}